\documentstyle[12pt,epsfig]{article}

\advance\textheight by \topskip
\begin{document}
\tolerance=100000
\thispagestyle{empty}
\setcounter{page}{0}

\newcommand{\be}{\begin{equation}}
\newcommand{\ee}{\end{equation}}
\newcommand{\br}{\begin{eqnarray}}
\newcommand{\er}{\end{eqnarray}}
\newcommand{\ba}{\begin{array}}
\newcommand{\ea}{\end{array}}
\newcommand{\bi}{\begin{itemize}}
\newcommand{\ei}{\end{itemize}}
\newcommand{\bn}{\begin{enumerate}}
\newcommand{\en}{\end{enumerate}}
\newcommand{\bc}{\begin{center}}
\newcommand{\ec}{\end{center}}
\newcommand{\ul}{\underline}
\newcommand{\ol}{\overline}
\newcommand{\eennH}{$e^+e^-\rightarrow  \bar\nu_e\nu_e H$}
\newcommand{\eennbb}{$e^+e^-\rightarrow  \bar\nu_e\nu_e b\bar b$}
\newcommand{\eennbbph}{$e^+e^-\rightarrow  \bar\nu_e\nu_e b\bar b\gamma$}
\newcommand{\Hbb}{$H \rightarrow  b\bar b$}
\newcommand{\Hbbph}{$H\rightarrow  b\bar b\gamma$}
\newcommand{\eennjjjj}{$e^+e^-\rightarrow \bar\nu_e\nu_e \mathrm{jjjj}$}
\newcommand{\eennjjjjph}{$e^+e^-\rightarrow\bar\nu_e\nu_e\mathrm{jjjj}\gamma$}
\newcommand{\Hjjjj}{$H\rightarrow \bar\nu_e\nu_e \mathrm{jjjj}$}
\newcommand{\Hjjjjph}{$H\rightarrow \bar\nu_e\nu_e \mathrm{jjjj}\gamma$}
\newcommand{\eeeeH}{$e^+e^-\rightarrow  e^+e^- H$}
\newcommand{\eeeebb}{$e^+e^-\rightarrow  e^+e^- b\bar b$}
\newcommand{\eeeebbph}{$e^+e^-\rightarrow  e^+e^- b\bar b\gamma$}
\newcommand{\eeeejjjj}{$e^+e^-\rightarrow e^+e^- \mathrm{jjjj}$}
\newcommand{\eeeejjjjph}{$e^+e^-\rightarrow e^+e^- \mathrm{jjjj}\gamma$}
\newcommand{\eezh}{$e^+e^-\rightarrow ZH$}
\newcommand{\uub}{$ u\bar u$}
\newcommand{\ddb}{$ d\bar d$}
\newcommand{\ssb}{$ s\bar s$}
\newcommand{\ccb}{$ c\bar c$}
\newcommand{\bbb}{$ b\bar b$}
\newcommand{\ttb}{$ t\bar t$}
\newcommand{\eeb}{$ e^+ e^-$}
\newcommand{\mumub}{$ \mu^+\mu^-$}
\newcommand{\tautaub}{$ \tau^+\tau^-$}
\newcommand{\veveb}{$ \bar\nu_e\nu_e$}
\newcommand{\vmvmb}{$ \bar\nu_\mu\nu_\mu $}
\newcommand{\vtvtb}{$ \bar\nu_\tau\nu_\tau $}
\newcommand{\lra}{\leftrightarrow}
\newcommand{\ar}{\rightarrow}
\newcommand{\sm}{${\cal {SM}}$}
\newcommand{\MH}{M_{H}}
\newcommand{\MW}{M_{W}}
\newcommand{\MZ}{M_{Z}}
\newcommand{\Dir}{\kern -6.4pt\Big{/}}
\newcommand{\Dirin}{\kern -10.4pt\Big{/}\kern 4.4pt}
\newcommand{\DDir}{\kern -7.6pt\Big{/}}
\newcommand{\DGir}{\kern -6.0pt\Big{/}}
\def\Ord{\buildrel{\scriptscriptstyle <}\over{\scriptscriptstyle\sim}}
\def\OOrd{\buildrel{\scriptscriptstyle >}\over{\scriptscriptstyle\sim}}
\def\pl #1 #2 #3 {{\it Phys.~Lett.} {\bf#1} (#2) #3}
\def\np #1 #2 #3 {{\it Nucl.~Phys.} {\bf#1} (#2) #3}
\def\zp #1 #2 #3 {{\it Z.~Phys.} {\bf#1} (#2) #3}
\def\pr #1 #2 #3 {{\it Phys.~Rev.} {\bf#1} (#2) #3}
\def\prep #1 #2 #3 {{\it Phys.~Rep.} {\bf#1} (#2) #3}
\def\jp #1 #2 #3 {{\it J.~Phys.} {\bf#1} (#2) #3}
\def\prl #1 #2 #3 {{\it Phys.~Rev.~Lett.} {\bf#1} (#2) #3}
\def\mpl #1 #2 #3 {{\it Mod.~Phys.~Lett.} {\bf#1} (#2) #3}
\def\rmp #1 #2 #3 {{\it Rev. Mod. Phys.} {\bf#1} (#2) #3}
\def\xx #1 #2 #3 {{\bf#1}, (#2) #3}
\def\preprint{{\it preprint}}

\begin{flushright}
{\large DFTT 47/96}\\ 
{\large Cavendish--HEP--96/11}\\ 
{\rm July 1996\hspace*{.5 truecm}}\\ 
\end{flushright}

\vspace*{\fill}

\begin{center}
{\Large \bf Higgs signals and hard photons\\ 
at the Next Linear Collider:\\
the ${ZZ}$-fusion channel in the Standard Model}\\[0.95cm]
{\large Stefano Moretti\footnote{E-mails: Moretti@to.infn.it; 
Moretti@hep.phy.cam.ac.uk.}}\\[0.25cm]
{\it Dipartimento di Fisica Teorica, Universit\`a di Torino,}\\
{\it and I.N.F.N., Sezione di Torino,}\\
{\it Via Pietro Giuria 1, 10125 Torino, Italy.}\\[0.25cm]
{\it Cavendish Laboratory, 
University of Cambridge,}\\ 
{\it Madingley Road,
Cambridge, CB3 0HE, United Kingdom.}\\[0.2cm]
\end{center}
\centerline{PACS numbers: 14.80.Bn, 14.70.Bh, 14.70.Hp, 29.17.+w.}
\vspace*{\fill}

\begin{abstract}
{\small
\noindent
In this paper, we extend the analyses carried out in a previous
article for $WW$-fusion to the case of Higgs production
via $ZZ$-fusion within the Standard Model at the Next Linear
Collider, in presence of electromagnetic radiation due real photon
emission.
Calculations are carried out at tree-level and rates 
of the leading order (LO) processes 
$e^+e^-\rightarrow e^+e^- H \ar e^+e^- b\bar b $ and
$e^+e^-\rightarrow e^+e^- H \ar e^+e^- WW 
\ar e^+e^- \mathrm{jjjj}$ are compared to those
of the next-to-leading order (NLO) reactions
$e^+e^-\rightarrow e^+e^- H (\gamma)\ar e^+e^- b\bar b \gamma$
and $e^+e^-\rightarrow e^+e^- H (\gamma)\ar e^+e^- WW (\gamma)
\ar e^+e^- \mathrm{jjjj}\gamma$, 
in the case of energetic and isolated photons.
}
\end{abstract}

\vspace*{\fill}
\newpage
\noindent
In a previous paper \cite{WWmio} we studied the effects of hard photon
emission in the $WW$-fusion reactions
\be\label{oldbbph}
e^+e^-\rightarrow\bar\nu_e\nu_e H (\gamma)\ar \bar\nu_e\nu_e b\bar b \gamma,
\ee
\be\label{oldWWph}
e^+e^-\rightarrow\bar\nu_e\nu_e H (\gamma)\ar \bar\nu_e\nu_e WW (\gamma)
\ar \bar\nu_e\nu_e \mathrm{jjjj}\gamma.
\ee
The production mechanism $e^+e^-\ar \bar\nu_e\nu_e H$ and the decay channels
$H\ar b\bar b,~\mathrm{jjjj}$  
represent some of the most likely signatures of
Higgs processes at the Next Linear Collider (NLC) \cite{epem2}--\cite{epem5}, 
within the Standard Model
(\sm), for a $H$ scalar in the intermediate (1) 
and heavy (2) mass range (IMR and HMR,
respectively)\footnote{For a NLC with $\sqrt s\Ord500$ GeV and 
$\MH\Ord2\MW$ the production rates of the
bremsstrahlung process $e^+e^-\ar ZH$ are comparable or
larger than those of the $WW$-fusion channel \cite{bremSM,fusionSM}.}. 

In that study we focused our attention to the case of hard and isolated
photons.
The motivation of that analysis came from the observation that
the r\^ole of energetic photons from the initial state
is expected to be determinant at the
$e^+e^-$ colliders of the next generation. In particular, whereas at 
LEP1, SLC and partially LEP2, the finite width of the gauge
boson resonances heavily suppresses events with
hard photons produced in the $e^+e^-$ annihilation subprocess, at a high 
energy NLC (when $\sqrt s\OOrd300$ GeV) this is no longer true. Furthermore,
since the energy of the beams is much larger, the probability 
that the incoming electrons and positrons radiate is somewhat 
bigger. Therefore, one naturally expects the NLC to produce 
rather copiously events
accompanied by hard electromagnetic (EM) emission and these cannot
be ignored in realistic phenomenological analyses.

In many instances, one can give account of the Initial State Radiation
(i.e., bremsstrahlung of photons from the incoming electron/positron lines)
by means of the so-called `electron structure functions' \cite{structure}.
In addition, for certain designs of the NLC, the effects due
to such a radiation are dominant with respect to those due
to beamsstrahlung and Linac energy spread phenomena \cite{ISR}:
such that, in phenomenological studies, one can consistently deal
with bremsstrahlung photons only.
However, on the one hand,
this approach is realistically applicable only in the case 
of {\sl annihilation}- and {\sl conversion}-type
Feynman diagrams: that is, when the $e^\pm$-lines
are connected to each other via $s$- or $t,u$-channel interactions
\cite{ISRcomplete}.
When the electron and positron lines are disconnected by gauge boson
currents, like in the case of $WW$-fusion processes,
a separation of the radiation emitted by the incoming fermion lines from that
radiated by the virtual $W$-boson lines (see Fig.~1--2 in Ref.~\cite{WWmio})
is indeed not gauge invariant. On the other hand, in experimental
data, radiative events with photons coming from the Higgs
production mechanism
are not distinguishable from those in which they come from the Higgs decays.

To give reliable predictions for the corrections at the order
${\cal O}(\alpha_{\mathrm{em}})$ to the leading
processes
\be\label{oldbb}
e^+e^-\rightarrow\bar\nu_e\nu_e H \ar \bar\nu_e\nu_e b\bar b,
\ee
\be\label{oldWW}
e^+e^-\rightarrow\bar\nu_e\nu_e H \ar \bar\nu_e\nu_e WW
\ar \bar\nu_e\nu_e \mathrm{jjjj},
\ee
one should compute the contributions due to real
photons as well as
those due to virtual photons in the loop diagrams.
Their summation through the orders ${\cal O}(\alpha_{\mathrm{em}}^5)$
and  ${\cal O}(\alpha_{\mathrm{em}}^7)$ would allow one
to cancel the infrared (soft and collinear) divergences appearing
in the matrix elements (MEs). However, to compare the tree-level rates
of the LO processes (3)--(4)
with those of the NLO ones (1)--(2) for hard photons only, allows one to assess
whether (at higher order) complications
should be expected in working out the position of the resonances and/or in
establishing their line-shape,
or whether the overall effect of the complete ${\cal O}(\alpha_{\mathrm{em}})$
corrections will be in the end only matter of a different normalisation (of the
LO spectra).

In the approach of Ref.~\cite{WWmio}, virtual photon contributions 
were ignored, for both the cases of what we called  `production radiation' 
and `decay radiation'. (In detail, the first is due to photons emitted 
during the Higgs production mechanism,
whereas
the second refers to photons radiated in the Higgs decays, and they
are separately gauge-invariant.) As loop diagrams were missing,
the mentioned summation was not performed there. In that paper,
we compared the rates of the lowest order processes 
(\ref{oldbb}) and (\ref{oldWW}) to those at next-to-lowest order
from reactions (\ref{oldbbph}) and (\ref{oldWWph}) over the phase space
regions defined by the cuts:
$p_T^{b,~\mathrm{j},\gamma}>1$ GeV and 
$\cos\theta_{b\gamma,~\mathrm{j}\gamma}<0.95$.
The aim was to make the point that in many instances (that is for various
combinations of NLC centre-of-mass (CM) 
energies, $\sqrt s$, and Higgs masses, $\MH$)
the rates of the hard radiative processes can be rather large compared to
those of the leading reactions and that in some cases 
a complete ${\cal O}(\alpha_{\mathrm{em}})$ calculation would be 
desirable in order to assess the relative importance of 
events of the type (\ref{WWph}) with hard photons, with respect to those in
which the electromagnetic emission is either infrared or virtual and 
to
the lowest order rates as well, as a shift in the position
of the Higgs peaks could be expected at NLO (see again Ref.~\cite{WWmio}
for a fuller discussion).
Another feature of interest to phenomenological analyses stressed in 
Ref.~\cite{WWmio}
is the fact that rather clear Higgs signals can appear 
in the spectra of the invariant masses $M_{b\bar b\gamma}$ and
$M_{\mathrm{jjjj}\gamma}$.
The last point outlined in Ref.~\cite{WWmio} is that it is generally
impossible to separate in radiative events (\ref{oldbbph})--(\ref{oldWWph}) the
`production radiation' from the `decay radiation', at least by exploiting
the spectra in photon energy, $E_\gamma$, and in
missing transverse momentum of the neutrino pair, $p_T^{\mathrm{miss}}$.

We intend in this brief report to perform studies similar to those of
Ref.~\cite{WWmio} for
the case of a \sm\ Higgs
boson produced via the $ZZ$-fusion mechanism and decaying through
the two mentioned channels, by studying the reactions:
\be\label{bbph}
e^+e^-\rightarrow e^+e^- H (\gamma)\ar e^+e^- b\bar b \gamma,
\ee
\be\label{WWph}
e^+e^-\rightarrow e^+e^- H (\gamma)\ar e^+e^- WW (\gamma)
\ar e^+e^- \mathrm{jjjj}\gamma,
\ee
\be\label{bb}
e^+e^-\rightarrow e^+e^- H \ar e^+e^- b\bar b,
\ee
\be\label{WW}
e^+e^-\rightarrow e^+e^- H \ar e^+e^- WW
\ar e^+e^- \mathrm{jjjj}.
\ee
As in this case infrared divergences can occur also in the bremsstrahlung
of soft and collinear photons off the $e^\pm$-lines in the 
final state, we impose that 
$\cos\theta_{e/b\gamma,e/~\mathrm{j}\gamma}<0.95$ (that is, separation
of the photon also from the outgoing electron/positron). Moreover, since
the $e^\pm$-particles in the final states are detectable 
(contrary to the neutrinos in the $WW$-fusion), we ask 
$p_T^{e,b,~\mathrm{j},\gamma}>1$ GeV.
The Feynman diagrams describing the reactions  (\ref{bbph})--(\ref{WWph})
at tree-level are the same pictured in Fig.~1--2 of Ref.~\cite{WWmio},
with graphs 5--6 there replaced by those in Figs.~1a--b of the present
paper. Their calculation has been carried out
in the same way as in Ref.~\cite{WWmio}, so we refer to that publication
for all details. 

We have verified that also in the case of the $ZZ$-fusion processes
the spectra in photon energy and in missing transverse momentum of 
the $e^+e^-$ pair show a similar behaviour to that described
in case of processes (\ref{oldbbph})--(\ref{oldWW}).
Therefore, we concentrate here only on the case of the integrated
rates and in the invariant mass spectra $M_{b\bar b}$, $M_{\mathrm{jjjj}}$, 
$M_{b\bar b\gamma}$ and $M_{\mathrm{jjjj}\gamma}$.
Our results are presented in Figs.~2a--b and Figs.~3a--b.
For consistency, the choice of the collider CM energies 
and Higgs masses we have adopted
here is the same as in Ref.~\cite{WWmio}.

Many of the features typical of processes (\ref{oldbbph})--(\ref{oldWW})
are reproduced in case of processes (\ref{bbph})--(\ref{WW})
as well: yet, differential and integrated rates are here 
smaller by an order of magnitude at least (compare Figs.~2a--b to 
Figs.~3--4 of Ref.~\cite{WWmio}), both at LO (as well known) and at NLO.
Concerning the total cross sections, one can notice (Fig.~2a) that in the 
IMR the 
NLO contribution due to hard photons is never greater than $\approx10\%$
of the LO rates, for all $\sqrt s=300,500$ and 1000 GeV,
when $\MH$ varies in the IMR. The same can be affirmed for 
a scalar with mass in the whole of the HMR (Fig.~2b).
 
Figs.~3a--b show the distribution in the invariant masses of the Higgs
decay products: i.e., $M_{b\bar b(\gamma)}$ and $M_{\mathrm{jjjj}(\gamma)}$.
Things are here very similar to the case
of the corresponding $WW$-fusion processes, such that the same
comments contained in Ref.~\cite{WWmio} hold here too. 
However, a different feature
of $ZZ$-fusion processes compared to the $WW$-fusion ones is
that the relative contribution of graphs 5 and 6 to the total
cross section at NLO is smaller in the former.
This is due to the additional cut between the final state electron/positron 
and photon directions implemented in case of processes 
(\ref{bbph})--(\ref{WWph}),
as the EM emission along the $e^\pm$ lines tends to be highly collinear.
This ultimately reduces the relevance of NLO hard photon 
events, compared to the lowest order rates. In fact,
for the $b\bar b$ invariant masses (left plots in Fig.~3a),
the smearing towards low values (that is, $M_{b\bar b}<\MH$) 
of the Higgs resonances due to the hard EM radiation
in the $H\ar b\bar b\gamma$ three-body decays induces an effect of only
$\approx1\%$ at the most, for all Higgs masses and collider
energies. The suppression of the `decay radiation' is once again visible
also in the
invariant mass of the $b\bar b\gamma$ system (plots on the right
in Fig.~3a), because of the absence of clear Breit-Wigner
peaks \cite{WWmio}. 
For a Higgs scalar in the HMR,
we notice that
the shape of the mass distributions on the left hand side of Fig.~3b
is rather different at the two orders. The effect is visible at all
energies: for example, see the open and shaded
dotted histograms (see also Ref.~\cite{WWmio}). However, the quantitative
effect is rather small compared to the case of $WW$-fusion.
Finally, 
Higgs peaks clearly appear in the $M_{\mathrm{jjjj}\gamma}$ spectra (right
hand side of Fig.~3b).

Therefore, we conclude by remarking that 
the corrections due to hard photons in reactions induced
by $ZZ$-fusion are generally smaller 
compared to the case of $WW$-fusion processes, both in the intermediate
and heavy mass range of $\MH$.
In particular then, when folding the real radiation computed here with the 
virtual one and with the lowest order rates to produce the complete
result at the ${\cal O}(\alpha_{\mathrm{em}})$ higher order, the smearing
of the resonant Higgs peaks should be less relevant than in the $WW$-fusion 
case,
and negligible in general.
However, in the HMR and especially for very large NLC
energies (i.e., around and above 1000 GeV, see Fig.~3b bottom left),
a complete NLO calculation would be desirable for phenomenological
analyses that consider $ZZ$-fusion on its own.
We finally notice that for $ZZ$-fusion Higgs processes (like for $WW$-fusion)
the inclusion of four-jet decays via the $ZZ$-channel (i.e., $H\ar ZZ\ar
\mathrm{jjjj}(\gamma)$) should reduce the relevance of hard radiative
contributions;
and that a careful treatment of the Higgs-strahlung process $e^+e^-\ar
ZH$ (followed by $Z\ar e^+e^-$) and of its interference with the 
$ZZ$-fusion mechanism would be needed (see the discussions
in the last two paragraphs of Section 
3 in Ref.~\cite{WWmio}). 
The last aspect is especially important when
dealing with processes (5)--(8), as the LO rates of the two 
Higgs mechanisms (i.e., $ZH$-production and $ZZ$-fusion)
are comparable for all combinations of $\sqrt s$ and $\MH$
considered here (whereas $WW$-fusion always dominates for $\sqrt s\OOrd500$ 
GeV). 

\vskip0.5cm
\noindent
We are grateful to the UK PPARC for support. 
This work was financed in part by the
Ministero dell' Universit\`a e della Ricerca Scientifica and  by the EC
Programme ``Human Capital and Mobility'', contract CHRX--CT--93--0357
(DG 12 COMA).






\subsection*{Figure Captions}

\begin{description}

\item[Fig.~1 ] New Feynman diagrams at tree-level which substitute
those corresponding to the $WW$-fusion channel when the photon is
emitted by $W$ bosons in the Higgs production
mechanism (see diagrams 5--6 in Figs.~1--2 of Ref.~\cite{WWmio}):
({\bf a}) in case of process (\ref{bbph});
({\bf b}) in case of process (\ref{WWph}).

\item[Fig.~2 ] Cross sections of the processes: 
({\bf a}) (\ref{bbph}) and (\ref{bb});
({\bf b}) (\ref{WWph}) and (\ref{WW}); as a function
of the Higgs mass, at 
$\sqrt s =300$ GeV (upper plot),
$\sqrt s =500$ GeV (central plot),
$\sqrt s =1000$ GeV (lower plot), 
after the cuts 
$p_T^{e,h,\gamma}>1$ GeV and $\cos\theta_{e/h\gamma}<0.95$, where 
$h=b$ or $\mathrm{j}$, as appropriate. 

\item[Fig.~3 ] Distributions in: 
({\bf a}) invariant mass of the $b\bar b$ (left plots) and 
                         of the $b\bar b\gamma$ (right plots) systems  
          for processes (\ref{bbph}) and (\ref{bb}),
          for $\MH=60,100,140$ GeV
          at $\sqrt s=300$$(500)$$[1000]$ GeV; 
({\bf b}) invariant mass of the $\mathrm{jjjj}$ (left plots) and 
                         of the $\mathrm{jjjj}\gamma$ (right plots) systems  
          for processes (\ref{WWph}) and (\ref{WW}),
          for $\MH=180,220,260$$(180,260,340)$$[180,300,420]$ GeV
          at $\sqrt s=300$$(500)$$[1000]$ GeV.
The following cuts have been applied:
$p_T^{e,h,\gamma}>1$ GeV and $\cos\theta_{e/h\gamma}<0.95$, where 
$h=b$ or $\mathrm{j}$, as appropriate. The values of the CM energy are:
$\sqrt s=300$ GeV (upper plots, bins of 2 GeV); 
$\sqrt s=500$ GeV (central plots, bins of 4 GeV); 
$\sqrt s=1000$ GeV (lower plots, bins of 5 GeV).
In the left plots the
upper histograms refer to rates from the non-radiative processes
(\ref{bb}) and (\ref{WW}), whereas the lower histograms correspond 
to rates from the radiative processes (\ref{bbph}) and (\ref{WWph}). 
The latter are shaded.

\end{description}
\vfill
\newpage

\begin{figure}[p]
~\epsfig{file=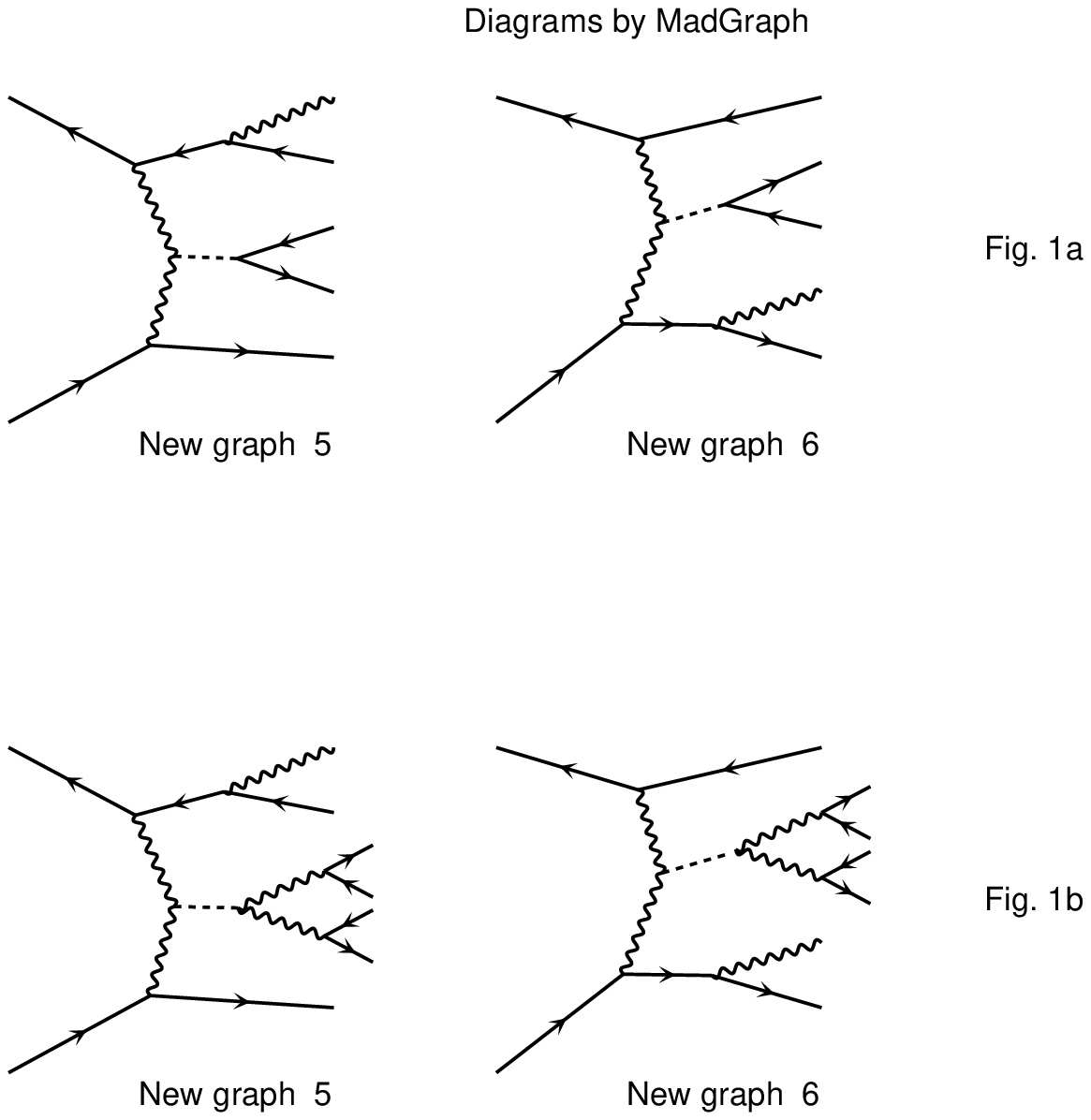,height=22cm}
\end{figure}
\stepcounter{figure}
\vfill
\clearpage

\begin{figure}[p]
~\epsfig{file=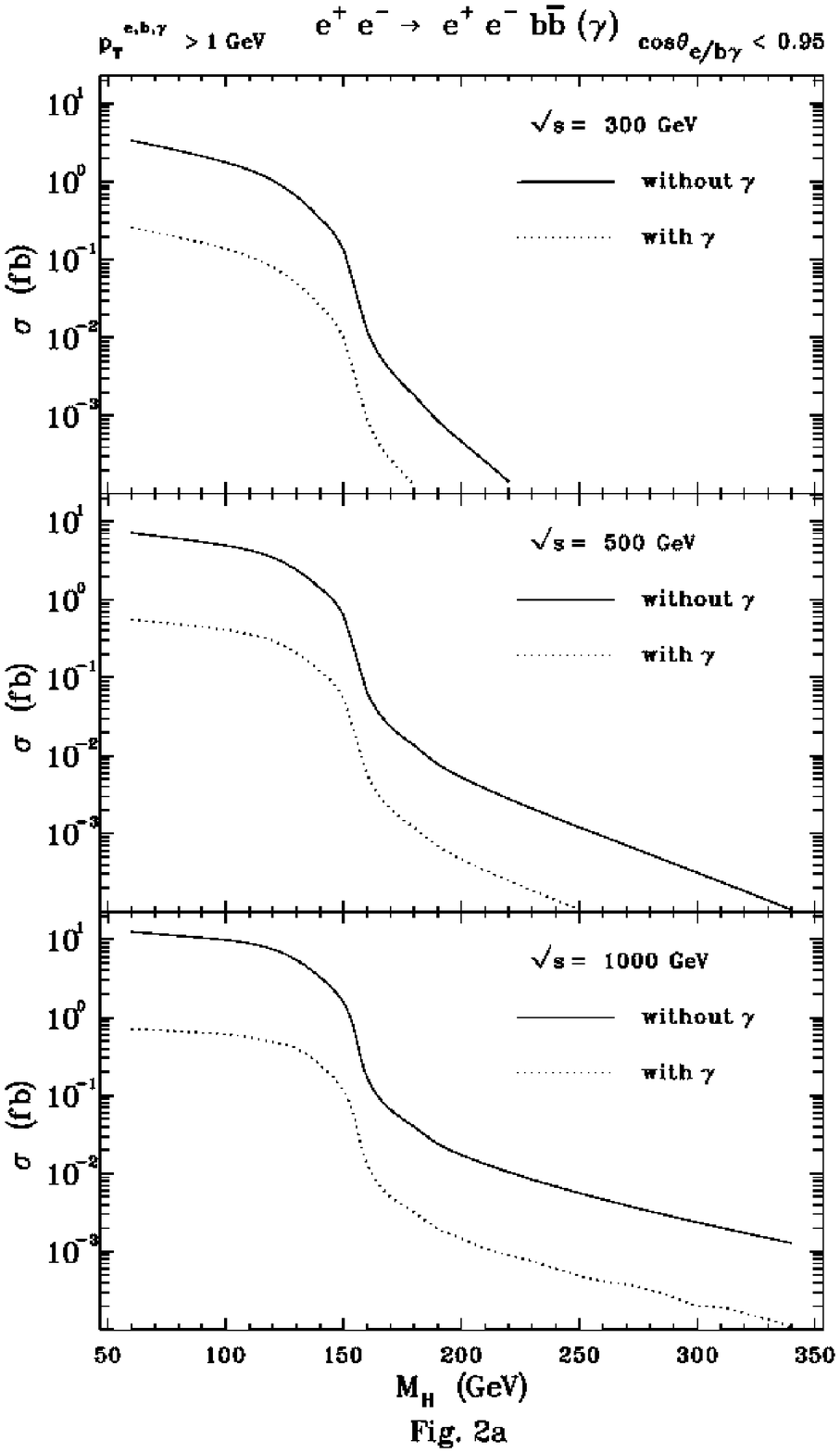,height=22cm}
\end{figure}
\stepcounter{figure}
\vfill
\clearpage

\begin{figure}[p]
~\epsfig{file=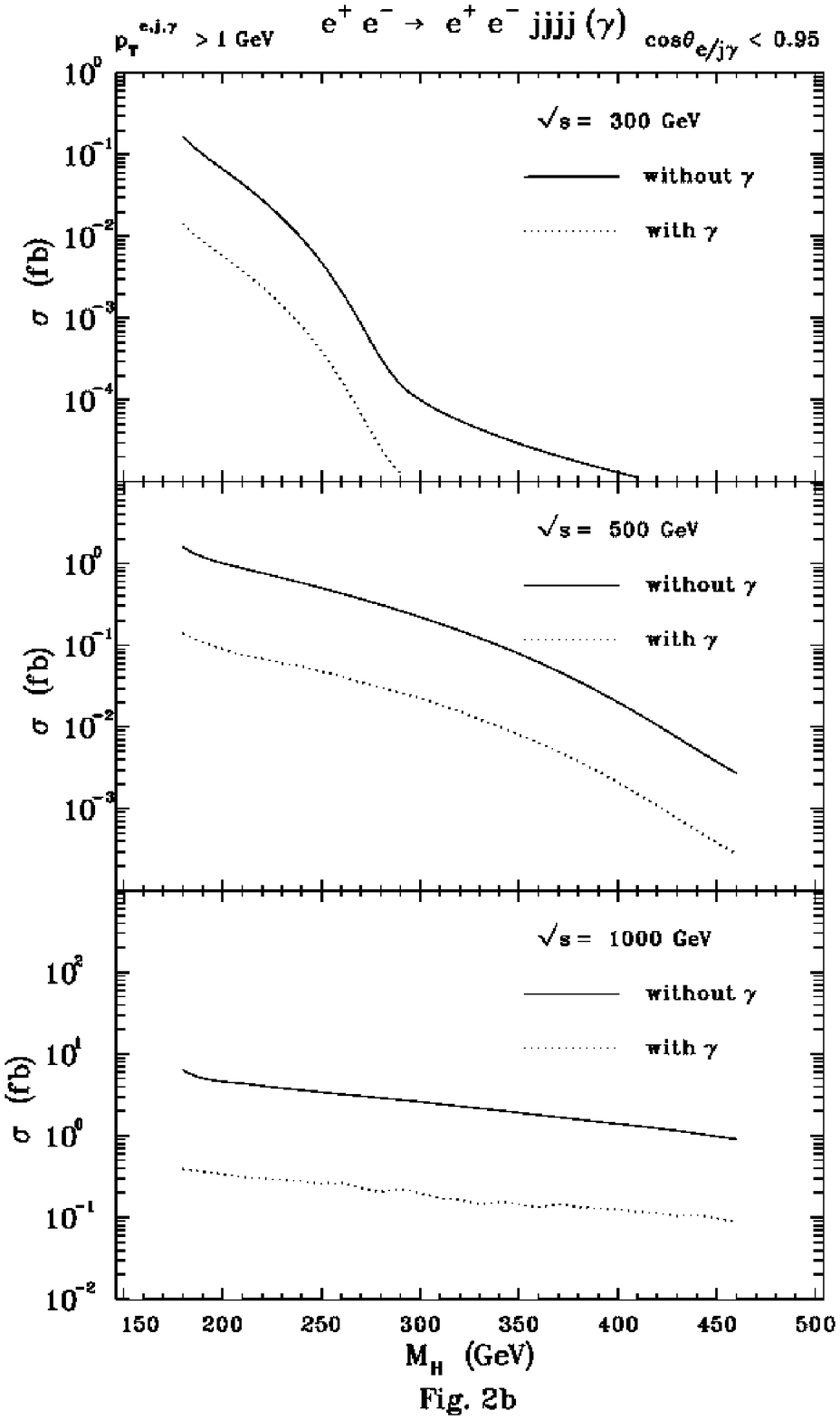,height=22cm}
\end{figure}
\stepcounter{figure}
\vfill
\clearpage

\begin{figure}[p]
~\epsfig{file=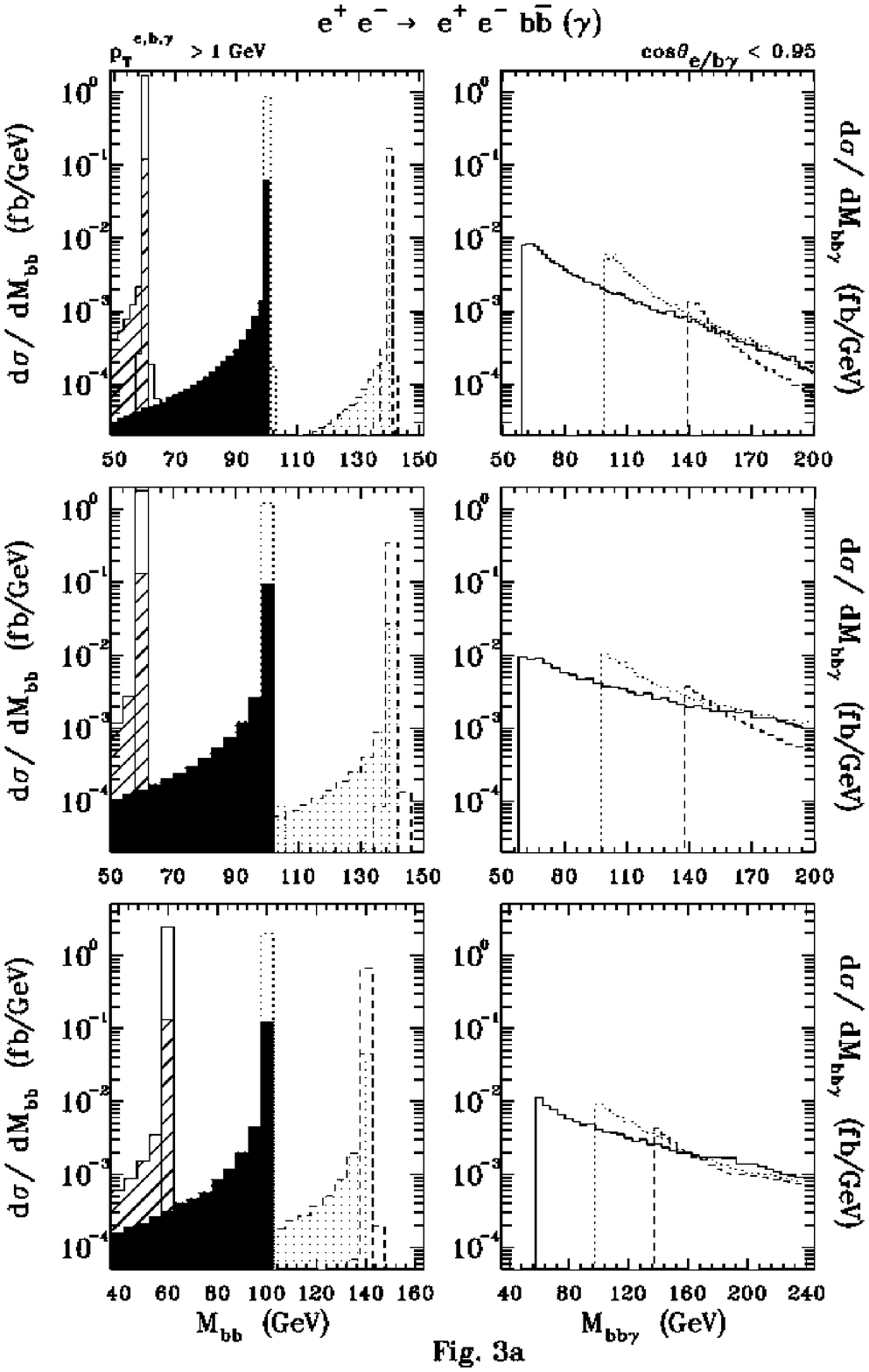,height=22cm}
\end{figure}
\stepcounter{figure}
\vfill
\clearpage

\begin{figure}[p]
~\epsfig{file=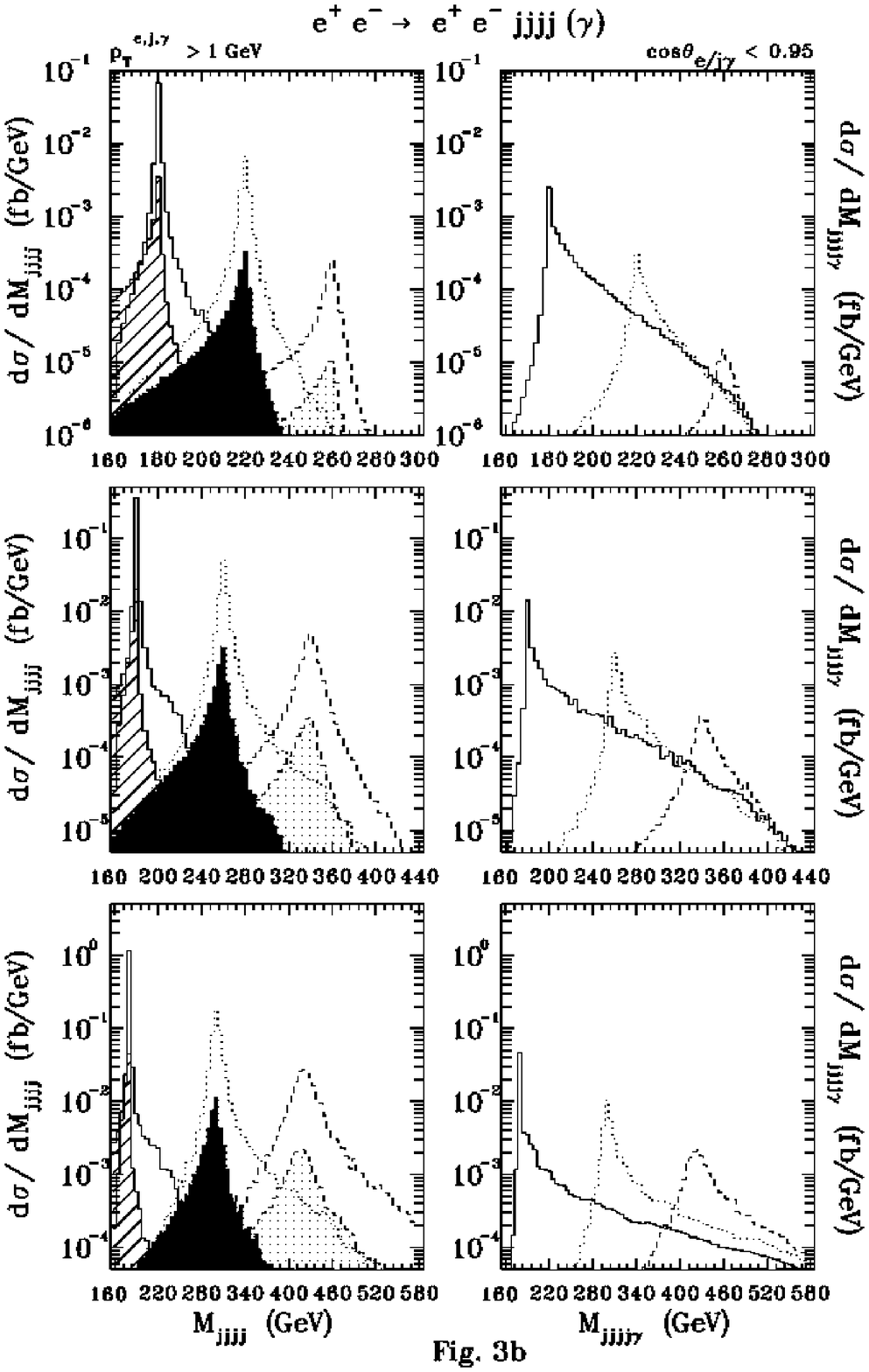,height=22cm}
\end{figure}
\stepcounter{figure}

\vfill
\end{document}